\documentclass{iopart}

\eqnobysec
\begin{document}
\title{Quasi-doubly periodic solutions to a generalized Lam\'e equation}
\author{Michael Pawellek}
\address{Institut fuer Theoretische Physik III, Universitaet Erlangen-Nuernberg, Staudtstr.7, D-91058 Erlangen, Germany}
\ead{michi@theorie3.physik.uni-erlangen.de}
\begin{abstract} 
We consider the generalized Lam\'e equation
\begin{eqnarray}
 \fl \frac{\mathrm{d}^2f}{\mathrm{d}x^2}+\frac{1}{2}\left(\frac{1}{x}+\frac{1}{x-1}+\frac{1}{x-k_1^{-2}}+\frac{1}{x-k_2^{-2}}\right)
 \frac{\mathrm{d}f}{\mathrm{d}x}- \nonumber\\ -\frac{Ek_1^{-2}k_2^{-2}+Ax^2-Bx}{4x(x-1)(x-k_1^{-1})(x-k_2^{-2})}f=0\nonumber
\end{eqnarray}
with \(A=\alpha+\beta k_1^{-2},\;B=\gamma k_2^{-2}+\delta k_1^{-2}+\lambda\).
By introducing a generalization of Jacobi's elliptic functions we transform this equation to a Schr\"odinger equation with
(quasi-doubly) periodic potential. We show that only for a finite set of integral values for the parameters 
\((\alpha,\beta,\gamma,\delta,\lambda)\) quasi-doubly periodic eigenfunctions expressible in terms of generalized Jacobi functions exist.
For this purpose we also establish a relation to the generalized Ince equation.

\end{abstract}
\pacs{02.30.Gp, 02.30.Hq, 03.65.Ge}

\section{Introduction}
It is well known \cite{Arsc,Erde,Turb} that the Lam\'e equation (in Jacobian form)
\begin{equation}\label{eq:Weierstrass}
 \frac{\mathrm{d}^2f}{\mathrm{dz^2}}-n(n+1)k^2\mathrm{sn}^2(z,k)f=-Ef
\end{equation}
for given \(n\in\mathbf{N}\) has \(2n+1\) doubly-periodic eigenfunctions which can be expressed as polynomials
in Jacobi elliptic functions \(\mathrm{sn}(z,k),\mathrm{cn}(z,k)\) and \(\mathrm{dn}(z,k)\). The Jacobian form of the Lam\'e equation
can be interpreted as a one-dimensional Schr\"odinger equation with periodic potential \(V(z)=-n(n+1)k^2\mathrm{sn}^2(z,k)\).
In algebraic form the Lam\'e equation is given by (where the substitution \(x=\mathrm{sn}^2(z,k)\) has been made in (\ref{eq:Weierstrass}))
\begin{equation}\label{eq:AlgebraicLame}
 \fl \frac{\mathrm{d}^2f}{\mathrm{d}x^2}+\frac{1}{2}\left(\frac{1}{x}+\frac{1}{x-1}+\frac{1}{x-k^{-2}}\right)
 \frac{\mathrm{d}f}{\mathrm{d}x}+\frac{Ek^{-2}-n(n+1)x}{4x(x-1)(x-k^{-2})}f=0
\end{equation}
and is of the Fuchsian type with four regular singular points \cite{Ince}. 

We consider the equation
\begin{eqnarray}\label{eq:GenLame}
 \fl \frac{\mathrm{d}^2f}{\mathrm{d}x^2}+\frac{1}{2}\left(\frac{1}{x}+\frac{1}{x-1}+\frac{1}{x-k_1^{-2}}+\frac{1}{x-k_2^{-2}}\right)
 \frac{\mathrm{d}f}{\mathrm{d}x}-\frac{Ek_1^{-2}k_2^{-2}+Ax^2-Bx}{4x(x-1)(x-k_1^{-2})(x-k_2^{-2})}f=0\nonumber\\
\end{eqnarray}
with \(A=\alpha+\beta k_1^{-2},\;B=\gamma k_2^{-2}+\delta k_1^{-2}+\lambda\) and \(0\leq k_2\leq k_1\leq 1\),
which is a generalization of the algebraic form of the Lam\'e equation (\ref{eq:AlgebraicLame}). It is of Fuchsian type with
five regular singular points. The exponents are \(0\) and \(1/2\) for \(x= 0, 1, k_1^{-2},k_2^{-2}\) and 
\(\frac{1}{2}[1\pm(1+\alpha+\beta k_1^{-2})^{1/2}]\)
for \(\infty\). This equation is of relevance when considering fluctuations 
around (anti-)periodic static solutions of (1+1)-dimensional scalar field theories with a \(\phi^6\) interaction term \cite{Muss}.

In the past several generalization of the original Lam\'e potential were considered, e.g. Darboux-Treibich-Verdier potentials \cite{Tre1, Gesz}, 
which can be written by a suitable variable transformation as Heun equations with four singular regular points \cite{Tak1}. Further
generalizations of Darboux-Treibich-Verdier potentials \cite{Tre2} can be written by a suitable variable transformations \cite{Tak2, Smir} as Fuchsian 
equations with more than four singular points. The additional singular points in the finite region of the complex plane are apparent \cite{Tak2}, which
means the exponents at these points are integers. (\ref{eq:GenLame}) has no apparent singular points in the finite region of the complex plane and is
therefore not covered by \cite{Tre2, Tak2, Smir}.

We show that (\ref{eq:GenLame}) can also be written as a Schr\"odinger equation with periodic potential. For this purpose we introduce a generalization of the 
Jacobi elliptic functions. These functions are quasi-doubly periodic but not elliptic in the strict sense, because of the appearance of cuts
in the complex plane. 

We will determine all values of the free parameters \((\alpha,\beta,\gamma,\delta,\lambda)\) for which quasi-doubly periodic eigenfunctions 
(with corresponding eigenvalue \(E\)) expressible as polynomials in terms of these generalized Jacobi functions exist. This can be done by transforming the
generalized Lam\'e equation to the generalized Ince equation (in \cite{Arsc,Magn} the Lam\'e equation has been transformed to the 
Ince equation in a similar way). The five-term recurrence relations obtained form the generalized Ince equation \cite{Reck} 
by inserting a Fourier ansatz enables us to get conditions for the existence of polynomial solutions.

The main result is that instead of infinitely many polynomial solutions, as in the case of the Lam\'e equation, the conditions
for existence of polynomial solutions in the generalized case are so restrictive that only for a finite set of values for 
the parameters \((\alpha,\beta,\gamma,\delta,\lambda)\) and \(E\) polynomial solutions exist.

\section{The generalized Jacobi functions}
In this section we introduce the generalized Jacobi functions and discuss some of their properties.
We consider the (pseudo-)hyperelliptic integral
\begin{equation}\label{eq:Hyperell}
 u(y,k_1,k_2)=\int_0^y\frac{\mathrm{d}t}{\sqrt{(1-t^2)(1-k_1^2t^2)(1-k_2^2t^2)}}
\end{equation}
where without loss of generality \(0<k_2<k_1<1\) are the moduli.
The inverse function \(y=s(u,k_1,k_2)\) fullfills the following
differential equation:
\begin{equation}\label{eq:Diff1}
 s'(u)^2=(1-s^2(u))(1-k_1^2s^2(u))(1-k_2^2s^2(u)).
\end{equation}
We define the companion functions for \(s(u)\) by (in most cases we use the abreviated notation \(s(u)\equiv s(u,k_1,k_2),
c(u)\equiv c(u,k_1,k_2)\), etc.)
\begin{equation}\label{eq:Companion}
 \fl c^2(u)=1-s^2(u),\qquad d_1^2(u)=1-k_1^2s^2(u),\qquad d_2^2(u)=1-k_2^2s^2(u).
\end{equation}
Without solving the integral (\ref{eq:Hyperell}) explicitly, one can derive certain properties of these functions. 
From (\ref{eq:Companion}) one gets
\begin{equation}\label{eq:Rel1}
 \fl d_i^2(u)-k_i^2c^2(u)=1-k_i^2,\;i=1,2;\qquad k_1^2d_2^2(u)-k_2^2d_1^2(u)=k_1^2-k_2^2.
\end{equation}
The first derivatives of these functions are given by
\begin{eqnarray}\label{eq:firstderiv}
 s'(u)=c(u)d_1(u)d_2(u),\qquad c'(u)=-s(u)d_1(u)d_2(u),\nonumber\\
 d_1'(u)=-k_1^2s(u)c(u)d_2(u),\qquad d_2'(u)=-k_2^2s(u)c(u)d_1(u),
\end{eqnarray}
which can easily be shown by applying (\ref{eq:Diff1}). The functions (\ref{eq:Companion}) with the properties (\ref{eq:firstderiv}) are generalizations 
of the usual Jacobi elliptic functions \(\mathrm{sn}(u),\mathrm{cn}(u)\) and \(\mathrm{dn}(u)\) with
\begin{equation}
 \fl \mathrm{sn}'(u)=\mathrm{cn}(u)\mathrm{dn}(u),\;\;\mathrm{cn}'(u)=-\mathrm{sn}(u)\mathrm{dn}(u),\;\;\mathrm{dn}'(u)=-k^2\mathrm{sn}(u)\mathrm{cn}(u)
\end{equation}
and they reduce to them for \(k_2\to 0\).

By using (\ref{eq:Companion}) the second derivatives can be written as
\begin{eqnarray}\label{eq:secderiv}
 \fl s''(u)=-3k_1^2k_2^2s^5(u)+2(k_1^2+k_2^2+k_1^2k_2^2)s^3(u)-(1+k_1^2+k_2^2)s(u)\\
 \fl c''(u)=-3k_1^2k_2^2c^5(u)-2(k_1^2+k_2^2-3k_1^2k_2^2)c^3(u)+(-1+2k_1^2+2k_2^2-3k_1^2k_2^2)c(u)\nonumber\\
 \fl d_1''(u)=-3k_2^2k_1^{-2}d_1^5(u)-2(1+k_2^2-3k_2^2k_1^{-2})d_1^3(u)+(2-k_1^2+2k_2^2-3k_2^2k_1^{-2})d_1(u).\nonumber
\end{eqnarray}

Normally the inversion of a single hyperelliptic integral is problematic \cite{Jaco}, to say the least. Historically this obstacle has lead to 
the development of algebraic geometry and the theory of theta functions \cite{Bake}. We need no 
sophisticated methods of algebraic geometry because (\ref{eq:Hyperell}) can be reduced to an elliptic integral by applying \(t=\sqrt{\tau}\) \cite{Byrd}. 
The functions \(s(u),c(u),d_1(u)\) and \(d_2(u)\) can then be expressed in terms of the standard Jacobi elliptic functions:
\begin{eqnarray}\label{eq:Solution}
 s(u,k_1,k_2)=\mathrm{sn}(k_2'u,\kappa)\left[1-k_2^2+k_2^2\mathrm{sn}^2(k_2'u,\kappa)\right]^{-1/2}\nonumber\\
 c(u,k_1,k_2)=k_2'\mathrm{cn}(k_2'u,\kappa)\left[1-k_2^2\mathrm{cn}^2(k_2'u,\kappa)\right]^{-1/2}\nonumber\\
 d_1(u,k_1,k_2)=(k_1^2-k_2^2)^{1/2}\mathrm{dn}(k_2'u,\kappa)\left[k_1^2-k_2^2\mathrm{dn}^2(k_2'u,\kappa)\right]^{-1/2}\nonumber\\
 d_2(u,k_1,k_2)=(k_1^2-k_2^2)^{1/2}\left[k_1^2-k_2^2\mathrm{dn}^2(k_2'u,\kappa)\right]^{-1/2}
\end{eqnarray}
with \(\kappa^2=(k_1^2-k_2^2)/(1-k_2^2)\), \(k_2'=\sqrt{1-k_2^2}\) and \(0\leq k_2\leq k_1\leq 1\). 
They have branch-cuts along \((u_1,u_2)\) and \((u_3,u_4)\) with 
\begin{eqnarray}
 u_1=i\frac{\mathrm{cn}^{-1}(k_2,\kappa')}{k_2'},\qquad u_2=-u_1+2i\frac{\mathbf{K}(\kappa')}{k_2'},\nonumber\\
 u_3=u_1+2\frac{\mathbf{K}(\kappa)}{k_2'}, \qquad u_4=u_2+2\frac{\mathbf{K}(\kappa)}{k_2'}
\end{eqnarray}
where \(\mathbf{K}(k)\) is the complete elliptic integral of the first kind and \(\kappa'=\sqrt{1-\kappa^2}\).
Now one can see that the elementary relations (\ref{eq:firstderiv}) are rather hidden when using the Jacobi representation (\ref{eq:Solution}).
It is also advantageous to use (\ref{eq:Companion}) to (\ref{eq:secderiv}) when working with these functions and not representation 
(\ref{eq:Solution}) together with the standard identities for Jacobi functions, which can be found in any textbook on elliptic functions 
\cite{Erde,Byrd}. With this set-up algebraic manipulations become very simply and straight-forward.. 

From the doubly-periodic properties of the Jacobian elliptic functions one can see that (\ref{eq:Solution}) are quasi-doubly periodic:
\begin{eqnarray}\label{eq:Doubly}
 s\left(u+\frac{4\mathbf{K}(\kappa)}{k_2'}\right)=s\left(u+\frac{2i\mathbf{K}(\kappa')}{k_2'}\right)=\pm s(u)\nonumber\\
 c\left(u+\frac{4\mathbf{K}(\kappa)}{k_2'}\right)=c\left(u+\frac{2\mathbf{K}(\kappa)+2i\mathbf{K}(\kappa')}{k_2'}\right)=\pm c(u)\nonumber\\
 d_1\left(u+\frac{2\mathbf{K}(\kappa)}{k_2'}\right)=d_2\left(u+\frac{4i\mathbf{K}(\kappa')}{k_2'}\right)=\pm d_1(u)\nonumber\\
 d_2\left(u+\frac{2\mathbf{K}(\kappa)}{k_2'}\right)=d_2\left(u+\frac{2i\mathbf{K}(\kappa')}{k_2'}\right)=\pm d_1(u)
\end{eqnarray}
Here an expression such as \(s(u+4\mathbf{K}(\kappa)k_2'^{-1})\) has to be interpreted as analytic continuation of \(s(u)\) along
a path from \(u\) to \(u+4\mathbf{K}(\kappa)k_2'^{-1}\). If the path avoids the cuts, the positive sign has to be chosen on the
right hand side of (\ref{eq:Doubly}). Choosing a path which crosses a cut one-time, one ends up with the negative sign.
So these functions are quasi-doubly periodic, depending on the path of analytic continuation.

\section{The generalized Lam\'e equation and its relation to the generalized Ince equation}
With the generalized Jacobi functions, defined in the last section, we can now introduce the "Jacobian" form of the generalized
Lam\'e equation (\ref{eq:GenLame}). This allows a discussion of this equation similar to the one done in \cite{Arsc,Magn} for the 
standard Lam\'e equation.

We refer to the 1-dimensional time-independent Schr\"odinger equation
\begin{equation}\label{eq:Schroed}
  \frac{\mathrm{d}^2f}{\mathrm{d}z^2}+V(z)f=-Ef
 \end{equation}
with the periodic potential
\begin{equation}\label{eq:Pot}
 \fl V(z)=(\alpha k_1^2k_2^2+\beta k_2^2)s^4(z,k_1,k_2)-(\gamma k_1^2+\delta k_2^2+\lambda k_1^2k_2^2)s^2(z,k_1,k_2).
\end{equation}
as generalized Lam\'e equation in Jacobian form, because by substitution of \(x=s^2(z,k_1,k_2)\)
(\ref{eq:Schroed}) transforms into (\ref{eq:GenLame}), which is a natural generalization of the algebraic form of the Lam\'e equation
(\ref{eq:AlgebraicLame}).
Also for \(k_2\to 0\) (\ref{eq:Pot}) reduces to the standard Lam\'e potential
\begin{equation}
 V(z)=-\gamma k_1^2\mathrm{sn}^2(z,k_1),
\end{equation}
which  for \(\gamma=n(n+1)\) with \(n\in\mathbf{N}\) has \(2n+1\) doubly-periodic solutions, the Lam\'e polynomials \cite{Arsc}.

By the substitution of \(t=a(z,k_1,k_2)\), where \(a(z,k_1,k_2)\) is defined by
\begin{equation}\label{eq:Substitution}
 \frac{\mathrm{d}t}{\mathrm{d}z}=\sqrt{(1-k_1^2\sin^2t)(1-k_2^2\sin^2t)}
\end{equation}
 (this can be understood as a generalization of Jacobi's amplitude function \(\mathrm{am}(z,k)\)) and using 
\begin{equation}
 \frac{\mathrm{d}^2t}{\mathrm{d}z^2}=\frac{1}{2}(k_1^2k_2^2-k_1^2-k_2^2)\sin(2t)-\frac{1}{4}k_1^2k_2^2\sin(4t),
\end{equation}
which follows directly from (\ref{eq:Substitution}), one can transform (\ref{eq:Schroed}) to the generalized Ince-equation \cite{Reck}
\begin{eqnarray}\label{eq:GenInce}
 \fl (1+a_1\cos(2t)+a_2\cos(4t))\frac{\mathrm{d}^2f}{\mathrm{d}t^2}+(b_1\sin(2t)+b_2\sin(4t))\frac{\mathrm{d}f}{\mathrm{d}t}+\nonumber\\
 +(c+d_1\cos(2t)+d_2\cos(4t))f=0
\end{eqnarray}
with coefficients
\begin{eqnarray}\label{eq:Koeff}
 a_1 =\frac{k_1^2+k_2^2-k_1^2k_2^2}{2-k_1^2-k_2^2+\frac{3}{4}k_1^2k_2^2},\qquad 
 a_2 =\frac{\frac{1}{4}k_1^2k_2^2}{2-k_1^2-k_2^2+\frac{3}{4}k_1^2k_2^2},\nonumber\\ 
 { } \nonumber\\
 b_1 =-a_1,\qquad b_2 =-2a_2,\nonumber\\
 { } \nonumber\\
 c=\frac{2E-\gamma k_1^2+(\frac{3\beta}{4}-\delta)k_2^2+(\frac{3\alpha}{4}-\lambda)k_1^2k_2^2}{2-k_1^2-k_2^2+\frac{3}{4}k_1^2k_2^2},\\
 { } \nonumber\\
\fl  d_1=\frac{\gamma k_1^2+(\delta-\beta)k_2^2+(\lambda-\alpha)k_1^2k_2^2}{2-k_1^2-k_2^2+\frac{3}{4}k_1^2k_2^2},\qquad
 d_2=\frac{\frac{1}{4}(\alpha k_1^2k_2^2+\beta k_2^2)}{2-k_1^2-k_2^2+\frac{3}{4}k_1^2k_2^2}.\nonumber
\end{eqnarray}

\subsection{Some remarks}
\begin{enumerate}
 \item The eigenvalue parameter \(E\) of (\ref{eq:Schroed}) only appears in the coefficient \(c\) of the generalized Ince equation.
 \item Solutions to (\ref{eq:Schroed}) with period \(2k_2'^{-1}\mathbf{K}(\kappa)\) and \(4k_2'^{-1}\mathbf{K}(\kappa)\) correspond
 to solutions to (\ref{eq:GenInce}) with period \(\pi\) and \(2\pi\) respectively.
\end{enumerate}

\section{The solutions}
In the following we find all values for \((\alpha,\beta,\gamma,\delta,\lambda)\) and \(E\), for which (\ref{eq:Schroed}) has polynomial solutions in
terms of (\ref{eq:Solution}). For this purpose it is advantageous to consider (\ref{eq:GenInce}). Because we are interested in periodic solutions, 
we can now make a Fourier expansion for the unknown solutions. Because (\ref{eq:GenInce}) has periodic coefficients with period \(\pi\), by 
Floquet's theorem \cite{Magn,Reck} it is sufficient to consider only solutions with 
period \(\pi\) or \(2\pi\). Therefore we have to consider four 
different Fourier expansions for the unknown solutions corresponding to even and odd functions with period \(\pi\) or \(2\pi\).

One ends up with five-term recurrence relations for the Fourier coefficients, which furnish conditions on the parameters 
\(\alpha,\beta,\gamma,\delta,\lambda\). In \cite{Reck} these recurrence relations were discussed in the context
of coexistence of two linearly independent periodic solutions to (\ref{eq:GenInce}).

\subsection{Even functions with period \(\pi\)}
Inserting the Fourier ansatz
\begin{equation}
 f(t)=\sum_{n=0}^{\infty}A_{2n}\cos(2nt)
\end{equation}
into (\ref{eq:GenInce}) gives the following reccurence relations \cite{Reck}:
\begin{eqnarray}
 -cA_0+Q_1(-1)A_2+Q_2(-2)A_4=0\nonumber\\
 Q_1(0)A_0+(4-c+Q_2(-1))A_2+Q_1(-2)A_4+Q_2(-3)A_6=0\nonumber\\
 \fl Q_2(n-2)A_{2n-4}+Q_1(n-1)A_{2n-2}+A_{2n}(4n^2-c)+\nonumber\\
 +Q_1(-n-1)A_{2n+2}+Q_2(-n-2)A_{2n+4}=0,\;n>1
\end{eqnarray}
or in matrix form
\begin{small}\begin{equation}\label{eq:Matrix1}
 \fl \left(\begin{array}{ccccccc}
  -c & Q_1(-1) & Q_2(-2) & 0 & 0 & 0 & \dots\\
  Q_1(0) & 4-c+Q_2(-1) & Q_1(-2) & Q_2(-3) & 0 & 0 & \dots\\
  Q_2(0) & Q_1(1) & 16-c & Q_1(-3) & Q_2(-4) & 0 & \dots\\
  0 & Q_2(1) & Q_1(2) & 36-c & Q_1(-4) & Q_2(-5) & \dots\\
  \vdots & & & \dots & & & \vdots\end{array}\right)\left(\begin{array}{c}
  A_0 \\ A_2 \\ A_4 \\ A_6 \\ \vdots\end{array}\right)=0
\end{equation}\end{small}
with
\begin{equation}
 Q_i(\mu)=2a_i\mu^2-b_i\mu-\frac{d_i}{2},\;i=1,2.
\end{equation}

(\ref{eq:GenInce}) has more than one polynomial solution with period \(\pi\) only if the following three equations have integral 
solutions for \(\mu\) \cite{Reck}:
\begin{equation}\label{eq:condition}
 Q_1(\mu)=Q_2(\mu)=Q_2(\mu-1)=0.
\end{equation}
If this is the case, the determinant of the infinite matrix in (\ref{eq:Matrix1}) separates into the product of determinants of a finite 
submatrix and an infinite matrix \cite{Reck}. From setting the determinant of the finite submatrix to zero one determines the allowed
values for \(c\) and from this the eigenvalues \(E\), see remark 3.1.(i). By the identity \(s(z,k_1,k_2)=\sin(a(z,k_1,k_2))\) 
polynomial solutions in \(\sin(t)\) for (\ref{eq:GenInce}) become polynomial solutions in \(s(z)\) for (\ref{eq:Schroed}).

From the second and third condition of (\ref{eq:condition}) follows the relation
\begin{equation}\label{eq:condition2}
 b_2=2a_2(2\mu-1),
\end{equation} 
and (\ref{eq:Koeff}) shows, that only \(\mu=0\) is permitted. On the other hand, in order that the first two 
conditions of (\ref{eq:condition}) are fullfilled by \(\mu=0\), one has to set \(d_1=d_2=0\). This is only possible, when 
\(\alpha=\beta=\gamma=\delta=\lambda=0\), see (\ref{eq:Koeff}). So (\ref{eq:GenInce}) cannot have more than one even polynomial solution 
with period \(\pi\) for any given values of \((\alpha,\beta,\gamma,\delta,\lambda)\).

In order that only one polynomial solution for given values of \((\alpha,\beta,\gamma,\delta,\lambda)\) exists,
it is necessary that only one column or row of the infinite matrix is zero. In the following we go through all possibilities, which give 
a nontrivial result.

\paragraph{First case.}
We set the first row to zero:
\begin{equation}\label{eq:firstrow}
 Q_1(-1)=Q_2(-2)=c=0.
\end{equation}
The first two equations of (\ref{eq:firstrow}) reduce to
\begin{eqnarray}
 (2-\gamma)k_1^2+(2-\delta+\beta)k_2^2+(\alpha-\lambda-2)k_1^2k_2^2=0\nonumber\\
 (8-\alpha)k_1^2k_2^2-\beta k_2^2=0.
\end{eqnarray}
These equations are only fullfilled for
\begin{equation}
 \alpha=8,\qquad \beta=0,\qquad \gamma=\delta=2,\qquad \lambda=6.
\end{equation}
The eigenvalue is determined by the third condition of (\ref{eq:firstrow}):
\begin{equation}
 E=k_1^2+k_2^2,
\end{equation}
and the eigenfunction is given by 
\begin{equation}
 f(z)=d_1(z)d_2(z)
\end{equation}
which can be checked by inspection.

\paragraph{Second case.}
We set the first column to zero:
\begin{equation}
 Q_1(0)=Q_2(0)=c=0.
\end{equation}
This reduces to
\begin{equation}
 d_1=d_2=0.
\end{equation}
So there is no nontrivial solution, see the discussion after (\ref{eq:condition2}).

\paragraph{Other cases.}
For the other columns one has the four conditions (in addition to \(4n^2-c=0\))
\begin{equation}
 Q_i(\pm\mu)=0,\;i=1,2.
\end{equation}
From \(Q_i(\mu)-Q_i(-\mu)=0\) it follows 
\begin{equation}
 2b_i\mu=0.
\end{equation}
This has only a nontrivial solution for \(\mu=0\), which is the second case considered just above.

For the other rows one must set \(Q_1(\mu)=Q_1(-\mu-2)=0\) and \(Q_2(\mu-1)=Q_2(-\mu-3)=0\). 
The first two conditions are only fullfilled for \(\mu=-1\), which is the first case considered above.

So there are no further solutions.

\subsection{Odd functions with period \(\pi\)}
Insertion of the Fourier ansatz
\begin{equation}
 f(t)=\sum_{n=0}^{\infty}B_{2n}\sin(2nt)
\end{equation}
in (\ref{eq:GenInce}) gives reccurence relations with the following matrix 
\begin{equation}
 \fl \left(\begin{array}{ccccccc}
  4-c -Q_2(-1) & Q_1(-2) & Q_2(-3) & 0 & 0 & 0 & \dots\\
  Q_1(1) & 16-c & Q_1(-3) & Q_2(-4) & 0 & 0 & \dots\\
  Q_2(1) & Q_1(2) & 36-c & Q_1(-4) & Q_2(-5) & 0 & \dots\\
  0 & Q_2(2) & Q_1(3) & 64-c & Q_1(-5) & Q_2(-6) & \dots\\
  \vdots & & & \dots & & & \vdots\end{array}\right).
\end{equation}
By the same arguments as in the last subsection more than one polynomial solution for given values of \((\alpha,\beta,\gamma,\delta,\lambda)\)
is not possible. The first column vanishes for
\begin{equation}
 \alpha=8,\qquad \beta=0,\qquad \gamma=\delta=6,\qquad \lambda=2.
\end{equation}
The corresponding eigenvalue and -function are given by
\begin{equation}
 E=4+k_1^2+k_2^2,\qquad f(z)=s(z)c(z).
\end{equation}
The first row vanishes for
\begin{equation}
 \alpha=24,\qquad \beta=0,\qquad \gamma=\delta=\lambda=12.
\end{equation}
The corresponding eigenvalue and -function are given by
\begin{equation}
 E=4(1+k_1^2+k_2^2),\qquad f(z)=s(z)c(z)d_1(z)d_2(z).
\end{equation}

\subsection{Odd functions with period \(2\pi\)}
Insertion of the Fourier ansatz
\begin{equation}
 f(t)=\sum_{n=0}^{\infty}B_{2n+1}\sin((2n+1)t)
\end{equation}
into (\ref{eq:GenInce}) gives recurrence relations with the following matrix
\begin{footnotesize}\begin{equation}\label{eq:Matrix3}
 \fl \left(\begin{array}{ccccccc}
 2-2c-Q_1^*(0) & Q_1^*(-1)-Q_2^*(-1) & Q_2^*(-2) & 0 & 0 & 0 & \dots\\
 Q_1^*(1)-Q_2^*(0) & 18-2c & Q_1^*(-2) & Q_2^*(-3) & 0 & 0 & \dots\\
 Q_2^*(1) & Q_1^*(2) & 50-2c & Q_1^*(-3) & Q_2^*(-4) & 0 & \dots\\
  0 & Q_2^*(2) & Q_1^*(3) & 96-2c & Q_1^*(-4) & Q_2^*(-5) & \dots\\
  \vdots & & & \dots & & & \vdots\end{array}\right).
\end{equation}\end{footnotesize}
with 
\begin{equation}
 Q_i^*(\mu)=a_i(2\mu-1)^2-b_i(2\mu-1)-d_i,\;i=1,2.
\end{equation}
 
(\ref{eq:GenInce}) has more than only one polynomial solution with period \(2\pi\) if the following three equations have integral 
solutions:
\begin{equation}
 Q_1^*(\mu)=Q_2^*(\mu)=Q_2^*(\mu-1)=0.
\end{equation}
From the third equation follows the relation
\begin{equation}
 b_2=4a_2(\mu-1).
\end{equation}
This cannot be fullfilled for our case, see (\ref{eq:Koeff}). More than one odd polynomial solution with period \(2\pi\) for
given values of \((\alpha,\beta,\gamma,\delta,\lambda)\) is not possible.

The solution for vanishing first column in (\ref{eq:Matrix3}) is given by
\begin{equation}
 \fl \alpha=3,\qquad \beta=0,\quad \gamma=\delta=\lambda=2,\qquad E=1+k_1^2+k_2^2,\qquad f(z)=s(z).
\end{equation}
The solution for vanishing first row in (\ref{eq:Matrix3}) is given by 
\begin{eqnarray}
 \fl \alpha=15,\qquad \beta=0,\qquad \gamma=\delta=6,\qquad \lambda=12, \qquad E=1+4(k_1^2+k_2^2),\nonumber \\ f(z)=s(z)d_1(z)d_2(z).
\end{eqnarray}

\subsection{Even functions with period \(2\pi\)}
Insertion of the Fourier ansatz
\begin{equation}
 f(t)=\sum_{n=0}^{\infty}A_{2n+1}\cos((2n+1)t)
\end{equation}
into (\ref{eq:GenInce}) gives recurrence relations with following matrix
\begin{footnotesize}\begin{equation}\label{eq:Matrix4}
 \fl \left(\begin{array}{ccccccc}
 2-2c-Q_1^*(0) & Q_1^*(-1)+Q_2^*(-1) & Q_2^*(-2) & 0 & 0 & 0 & \dots\\
 Q_1^*(1)+Q_2^*(0) & 18-2c & Q_1^*(-2) & Q_2^*(-3) & 0 & 0 & \dots\\
 Q_2^*(1) & Q_1^*(2) & 50-2c & Q_1^*(-3) & Q_2^*(-4) & 0 & \dots\\
  0 & Q_2^*(2) & Q_1^*(3) & 96-2c & Q_1^*(-4) & Q_2^*(-5) & \dots\\
  \vdots & & & \dots & & & \vdots\end{array}\right).
\end{equation}\end{footnotesize}
Also here, more than one even polynomial solution with period \(2\pi\) is not possible.

The solution for vanishing first column in (\ref{eq:Matrix4}) is given by
\begin{equation}
 \fl \alpha=3,\qquad\beta=0,\qquad\gamma=\delta=2,\qquad\lambda=0,\qquad E=1,\qquad f(z)=c(z).
\end{equation}

The solution for vanishing first row in (\ref{eq:Matrix4}) is given by
\begin{eqnarray}
 \alpha=15,\qquad\beta=0,\qquad\gamma=\delta=\lambda=6,\qquad E=1+k_1^2+k_2^2,\nonumber\\ f(z)=c(z)d_1(z)d_2(z).
\end{eqnarray}

\subsection{Further polynomial solutions}
The substitution \(f(z)=d_1(z)g(z)\) in (\ref{eq:Schroed}) yields
\begin{equation}\label{eq:transequ}
  \fl d_1(z)g''(z)+2d_1'(z)g'(z)+d_1''(z)g(z)+V(z)d_1(z)g(z)=-Ed_1(z)g(z).
\end{equation}
Applying the transformation defined by (\ref{eq:Substitution}) and using (\ref{eq:secderiv}), (\ref{eq:transequ}) can be transformed
to a generalized Ince equation (\ref{eq:GenInce}) for the unkown function \(g(t)\) with the following coefficients
\begin{eqnarray}
 a_1=\frac{k_1^2+k_2^2-k_1^2k_2^2}{2-k_1^2-k_2^2+\frac{3}{4}k_1^2k_2^2}, \qquad
 a_2=\frac{\frac{1}{4}k_1^2k_2^2}{2-k_1^2-k_2^2+\frac{3}{4}k_1^2k_2^2}\nonumber\\
 b_1=\frac{-3k_1^2-k_2^2+2k_1^2k_2^2}{2-k_1^2-k_2^2+\frac{3}{4}k_1^2k_2^2}, \qquad
 b_2=-4a_2\nonumber\\
 c=\frac{2E-\gamma k_1^2+(\frac{3}{4}\beta-\delta)k_2^2+(2-\lambda+\frac{3}{4}(\alpha-3))k_1^2k_2^2}{2-k_1^2-k_2^2+\frac{3}{4}k_1^2k_2^2}
  \nonumber\\
 d_1=\frac{(\gamma-2)k_1^2+(\delta-\beta)k_2^2+(\lambda-\alpha+1)k_1^2k_2^2}{2-k_1^2-k_2^2+\frac{3}{4}k_1^2k_2^2}\nonumber\\
 d_2=\frac{\frac{1}{4}\beta k_2^2+\frac{1}{4}(\alpha-3)k_1^2k_2^2}{2-k_1^2-k_2^2+\frac{3}{4}k_1^2k_2^2}.
\end{eqnarray}
Now one can perform the same steps as in sections 4.1 to 4.4. The additional (together with the previously obtained) solutions can be 
found in Table \ref{tab:list}. 

The substitution of \(f(z)\) by \(d_2(z)g(z)\) into (\ref{eq:Schroed}) and applying the same
steps as above only reproduces the previously obtained solutions.

\section{Expansion in \(s(z)\)}
An alternative way to find the previous solutions is to substitute a formal power series in generalized Jacobi functions e.g. 
\begin{equation}
 f(z)=\sum_{n=0}^{\infty}a_{2n+1}s^{2n+1}(z)
\end{equation}
into the generalized Lam\'e equation (\ref{eq:Schroed}) (a similar discussion of the Lam\'e equation can be found in
\cite{Arsc}). One gets four-term recurrence relations given in matrix form by
\begin{equation}\label{eq:recurrmat}
\left(\begin{array}{cccccc}
 D(0) & f(0) & 0 & 0 & 0 & \dots \\
 M_1(1) & D(1) & f(1) & 0 & 0 & \dots\\
 M_2(2) & M_1(2) & D(2) & f(2) & 0 & \dots\\
 0 & M_2(3) & M_1(3) & D(3) & f(3) & \dots\\
 \vdots & & &\dots & & \vdots\end{array}\right)
 \left(\begin{array}{c} a_1 \\ a_3 \\ a_5 \\ a_7 \\ \vdots\end{array}\right)=0
\end{equation}
with
\begin{eqnarray}
 D(n)=E-(2n+1)^2(1+k_1^2+k_2^2)\nonumber\\
 f(n)=2(n+1)(2n+3) \nonumber\\
 \fl M_1(n)=(2n(2n-1)-\gamma)k_1^2+(2n(2n-1)-\delta)k_2^2+(2n(2n-1)-\lambda)k_1^2k_2^2 \nonumber\\
 M_2(n)=(\alpha-(2n-3)(2n-1))k_1^2k_2^2+\beta k_2^2
\end{eqnarray}
The recurrence chain terminates if the following three equations have integral solutions for \(\mu\)
\begin{equation}
 M_1(\mu)=M_2(\mu)=M_2(\mu+1)=0.
\end{equation}
\(M_2(n_i)\) cannot be simultaneously zero for two different integral \(n_1\) and \(n_2\). So here one finds no
polynomial solution.

The recurrence chain also terminates if one row or column is zero. In (\ref{eq:recurrmat}) only for the first
column this can be done
\begin{equation}
 D(0)=M_1(1)=M_2(2)=0.
\end{equation}
This is the case for
\begin{equation}
 \alpha=3,\qquad\beta=0,\qquad\gamma=\delta=\lambda=2
\end{equation}
One finds again the solution \(f(z)=s(z)\) with eigenvalue \(E=1+k_1^2+k_2^2\). The recurrence relations for the other 
possible power series expansions can be found in the Appendix.

\section{Conclusion}  

\begin{table}
\caption{\label{tab:list}All 15 doubly-periodic solutions with corresponding eigenvalues and parameters.}
\begin{indented}
\item \normalsize\begin{tabular}{@{}cll}
\br
 \((\alpha,\beta,\gamma,\delta,\lambda)\) & eigenvalue & eigenfunction \\ \hline\hline
 (3,0,2,2,2)& \(E=1+k_1^2+k_2^2\) & \(f(z)=s(z)\) \\
 (3,0,2,2,0)& \(E=1\) & \(f(z)=c(z)\) \\
 (3,0,2,0,2)& \(E=k_1^2\) & \(f(z)=d_1(z) \) \\
 (3,0,0,2,2)& \(E=k_2^2\) & \(f(z)=d_2(z)\) \\
 (8,0,6,6,2)& \(E=4+k_1^2+k_2^2\) & \(f(z)=s(z)c(z)\)\\
 (8,0,2,2,6) & \(E=k_1^2+k_2^2\) & \(f(z)=d_1(z)d_2(z)\)\\
 (8,0,6,2,6) & \(E=1+4k_1^1+k_2^2\) & \(f(z)=s(z)d_1(z)\) \\
 (8,0,2,6,6) & \(E=1+k_1^2+4k_2^2\) & \(f(z)=s(z)d_2(z)\) \\
 (8,0,6,2,2) & \(E=1+k_1^2\) & \(f(z)=c(z)d_1(z)\) \\
 (8,0,2,6,2) & \(E=1+k_2^2\) & \(f(z)=c(z)d_2(z)\) \\
 (15,0,6,6,6)& \(E=1+k_1^2+k_2^2\) & \(f(z)=c(z)d_1(z)d_2(z)\) \\
 (15,0,12,6,6) & \(E=4+4k_1^2+k_2^2\) & \(f(z)=s(z)c(z)d_1(z)\) \\
 (15,0,6,12,6) & \(E=4+k_1^2+4k_2^2\) & \(f(z)=s(z)c(z)d_2(z)\) \\
 (15,0,6,6,12)& \(E=1+4(k_1^2+k_2^2)\) & \(f(z)=s(z)d_1(z)d_2(z)\) \\
 (24,0,12,12,12)& \(E=4(1+k_1^2+k_2^2)\) & \(f(z)=s(z)c(z)d_1(z)d_2(z)\) \\
\br
\end{tabular}
\end{indented}
\end{table}

We have shown that for the generalized Lam\'e equation (\ref{eq:Schroed}), which in algebraic form  (\ref{eq:GenLame}) is an ordinary 
differential equation of Fuchsian type with five regular singular points, only a finite number of quasi-doubly periodic solutions exist, which 
can be expressed in terms of generalized Jacobi functions (see Table \ref{tab:list}).

For this we have transformed the generalized Lam\'e equation to the generalized Ince equation and applied  
recently found  \cite{Reck} properties of their five-term recurrence relations. 
When (\ref{eq:Schroed}) is interpreted as Schr\"odinger equation every eigenfunction corresponds to a different periodic
potential (\ref{eq:Pot}) with certain values of the parameters (\(\alpha,\beta,\gamma,\delta,\lambda\)). So the generalized Lam\'e equation 
(\ref{eq:Schroed}) is not a quasi-exactly solvable differential equation in the sense of \cite{Turb}. Rather, it ranges between quasi-exactly solvable
and not exactly solvable differntial equations. Potentials where the solvabilty depends on the parameters are called "conditionally solvable potentials" 
\cite{Dutr} and were first observed in \cite{Fles}.

\appendix
\section{Recurrence relations for power series expansions in generalized Jacobi functions}
For completeness we list in this appendix all recurrence relations, which result from inserting several formal power series 
in \(s(z), c(z), d_1(z)\) and \(d_2(z)\) into the generalized Lam\'e equation (\ref{eq:Schroed}).
All these power series expansions of the unknown eigenfunctions of the generalized Lam\'e equation (\ref{eq:Schroed}) result in 
a recurrence matrix of the following structure
\begin{equation}
\left(\begin{array}{cccccc}
 D(0) & f(0) & 0 & 0 & 0 & \dots \\
 M_1(1) & D(1) & f(1) & 0  & 0 & \dots\\
 M_2(2) & M_1(2) & D(2) & f(2) & 0 & \dots\\
 0 & M_2(3) & M_1(3) & D(3) & f(3) & \dots\\
 \vdots & & \dots & & &\vdots\end{array}\right)
\end{equation}
which was discussed in Section 5. With this at hand one can also find in principle eigenfunctions of (\ref{eq:Schroed}) which
are only expressible as infinite series in the generalized Jacobi functions.
\subsection{ }
Insertion of the ansatz
\begin{equation}
 f(z)=\sum_{n=0}^{\infty}a_{2n}s^{2n}(z)
\end{equation}
into (\ref{eq:Schroed}) gives the following matrix elements
\begin{eqnarray}
 D(n)=E-4n^2(1+k_1^2+k_2^2) \nonumber\\
 f(n)=2(n+1)(2n+1) \nonumber\\
 M_1(n)=2(n-1)(2n-1)(k_1^2+k_2^2+k_1^2k_2^2)-(\gamma k_1^2+\delta k_2^2+\lambda k_1^2k_2^2) \nonumber\\
 M_2(n)=(\alpha-4(n-1)(n-2))k_1^2k_2^2+\beta k_2^2
\end{eqnarray}

\subsection{ }
The ansatz
\begin{equation}
 f(z)=c(z)\sum_{n=0}^{\infty} a_{2n}s^{2n}(z)
\end{equation}
gives a recurrence matrix with the following entries:
\begin{eqnarray}
 \fl D(n)=E-(2n+1)^2-4n^2(k_1^2+k_2^2) \nonumber\\
 \fl f(n)=2(n+1)(2n+1) \nonumber\\
 \fl M_1(n)=(2+2(n-1)(2n+1)-\gamma)k_1^2+(2+2(n-1)(2n+1)-\delta)k_2^2+\nonumber\\
 +(2(2n-1)(n-1)-\lambda)k_1^2k_2^2 \nonumber\\
 \fl M_2(n)=(\alpha-3-4n(n-2))k_1^2k_2^2+\beta k_2^2 
\end{eqnarray}
\subsection{ }
The ansatz
\begin{equation}
  f(z)=c(z)\sum_{n=0}^{\infty} a_{2n+1}s^{2n+1}(z)
\end{equation}
gives 
\begin{eqnarray}
 \fl D(n)=E-4(n+1)^2-(2n+1)^2(k_1^2+k_2^2) \nonumber\\
 \fl f(n)=2(n+1)(2n+3) \nonumber\\
 \fl M_1(n)=(2+2(2n-1)(n+1)-\gamma)k_1^2+(2+2(2n-1)(n+1)-\delta)k_2^2+\nonumber\\
 +(2n(2n-1)-\lambda)k_1^2k_2^2 \nonumber\\
 \fl M_2(n)=(\alpha-3-(2n-3)(2n+1))k_1^2k_2^2+\beta k_2^2
\end{eqnarray}

\subsection{ }
The ansatz 
\begin{equation}
 f(z)=d_1(z)\sum_{n=0}^{\infty} a_{2n}s^{2n}(z)
\end{equation}
gives
\begin{eqnarray}
 \fl D(n)=E-(2n+1)^2k_1^2-4n^2(1+k_2^2) \nonumber\\
 \fl f(n)=2(n+1)(2n+1) \nonumber\\
 \fl M_1(n)=(2+2(n-1)(2n+1)-\gamma)k_1^2+(2(n-1)(2n-1)-\delta)k_2^2+\nonumber\\
 +(2+2(n-1)(2n+1)-\lambda)k_1^2k_2^2 \nonumber\\
 \fl M_2(n)=(\alpha-3-4n(n-2))k_1^2k_1^2+\beta k_2^2
\end{eqnarray}

\subsection{ }
The ansatz
\begin{equation}
  f(z)=d_1(z)\sum_{n=0}^{\infty} a_{2n+1}s^{2n+1}(z)
\end{equation}
gives
\begin{eqnarray}
 \fl D(n)=E-4(n+1)^2k_1^2-(2n+1)^2(1+k_2^2) \nonumber\\
 \fl f(n)=2(n+1)(2n+3) \nonumber\\
 \fl M_1(n)=(2+2(2n-1)(n+1))k_1^2+(2n(2n-1)-\delta)k_2^2+\nonumber\\
 +(2+2(2n-1)(n+1)-\lambda)k_1^2k_2^2\nonumber\\
 \fl M_2(n)=(\alpha-3-(2n-3)(2n+1))k_1^2k_2^2+\beta k_2^2
\end{eqnarray}

\subsection{ }
The ansatz
\begin{equation}
 f(z)=c(z)d_1(z)\sum_{n=0}^{\infty} a_{2n}s^{2n}(z)
\end{equation}
gives
\begin{eqnarray}
 \fl D(n)=E-(2n+1)^2(1+k_1^2)-4n^2k_2^2 \nonumber\\
 \fl f(n)=2(n+1)(2n+1) \nonumber\\
 \fl M_1(n)=(6+2(n-1)(2n+3)-\gamma)k_1^2+(2+2(n-1)(2n+1)-\delta)k_2^2+\nonumber\\
 +(2+2(n-1)(2n+1)-\lambda)k_1^2k_2^2 \nonumber\\
 \fl M_2(n)=(\alpha-8-4(n-2)(n+2))k_1^2k_2^2+\beta k_2^2
\end{eqnarray}

\subsection{ }
The ansatz
\begin{equation}
  f(z)=c(z)d_1(z)\sum_{n=0}^{\infty} a_{2n+1}s^{2n+1}(z)
\end{equation}
gives
\begin{eqnarray}
 \fl D(n)=E-4(n+1)^2(1+k_1^2)-(2n+1)^2k_2^2 \nonumber\\
 \fl f(n)=2(n+1)(2n+3) \nonumber\\
 \fl M_1(n)=(6+2(2n-1)(n+2)-\gamma)k_1^2+(2+2(2n-1)(n+1)-\delta)k_2^2+\nonumber\\
 +(2+2(2n-1)(n+1)-\lambda)k_1^2k_2^2\nonumber\\
 \fl M_2(n)=(\alpha-8-(2n-3)(2n+3))k_1^2k_2^2+\beta k_2^2
\end{eqnarray}

\subsection{ }
The ansatz
\begin{equation}
 f(z)=d_1(z)d_2(z)\sum_{n=0}^{\infty} a_{2n}s^{2n}(z)
\end{equation}
gives
\begin{eqnarray}
 \fl D(n)=E-(2n+1)^2(k_1^2+k_2^2)-4n^2 \nonumber\\
 \fl f(n)=2(n+1)(2n+1) \nonumber\\
 \fl M_1(n)=(2+2(n-1)(2n+1)-\gamma)k_1^2+(2+2(n-1)(2n+1)-\delta)k_2^2+\nonumber\\
 +(6+2(n-1)(2n-3)-\lambda)k_1^2k_2^2\nonumber\\
 \fl M_2(n)=(\alpha-8-4n(n-2))k_1^2k_2^2+\beta k_2^2
\end{eqnarray}

\subsection{ }
The ansatz
\begin{equation}
  f(z)=d_1(z)d_2(z)\sum_{n=0}^{\infty} a_{2n+1}s^{2n+1}(z)
\end{equation}
gives
\begin{eqnarray}
 \fl D(n)=E-4(n+1)^2(k_1^2+k_2^2)-(2n+1)^2 \nonumber\\
 \fl f(n)=2(n+1)(2n+3) \nonumber\\
 \fl M_1(n)=(2+2(n+1)(2n-1)-\gamma)k_1^2+(2+2(n+1)(2n-1)-\delta)k_2^2+\nonumber\\
 +(6+2(n+2)(2n-1)-\lambda)k_1^2k_2^2\nonumber\\
 \fl M_2(n)=(\alpha-8+(2n-3)(2n+3))k_1^2k_2^2+\beta k_2^2
\end{eqnarray}

\subsection{ }
The ansatz
\begin{equation}
 f(z)=c(z)d_1(z)d_2(z)\sum_{n=0}^{\infty} a_{2n}s^{2n}(z)
\end{equation}
gives
\begin{eqnarray}
 \fl D(n)=E-(2n+1)^2(1+k_1^2+k_2^2)\nonumber\\
 \fl f(n)=2(n+1)(2n+1)\nonumber\\
 \fl M_1(n)=(6+2(n-1)(2n+3)-\gamma)k_1^2+(6+2(n-1)(2n+3)-\delta)k_2^2+\nonumber\\
 +(6+2(n-1)(2n+3)-\lambda)k_1^2k_2^2\nonumber\\
 \fl M_2(n)=(\alpha-15-4(n-2)(n+2))k_1^2k_2^2+\beta k_2^2
\end{eqnarray}

\subsection{ }
The ansatz
\begin{equation}
  f(z)=c(z)d_1(z)d_2(z)\sum_{n=0}^{\infty} a_{2n+1}s^{2n+1}(z)
\end{equation}
gives
\begin{eqnarray}
 \fl D(n)=E-4(n+1)^2(1+k_1^2+k_2^2) \nonumber\\
 \fl f(n)=2(n+1)(2n+3) \nonumber\\
 \fl M_1(n)=(6+2(2n-1)(n+2)-\gamma)k_1^2+(6+2(2n-1)(n+2)-\delta)k_2^2+\nonumber\\
 +(6+2(2n-1)(n+2)-\lambda)k_1^2k_2^2\nonumber\\
 \fl M_2(n)=(\alpha-15-(2n-3)(2n+5))k_1^2k_2^2+\beta k_2^2
\end{eqnarray}

\end{document}